# The 'Delayed Choice Quantum Eraser' Neither Erases Nor Delays

R. E. Kastner
University of Maryland, College Park, MD 20742

Abstract.
It is demonstrated that 'quantum eraser' (QE) experiments do not erase any information. Nor do they demonstrate 'temporal nonlocality' in their 'delayed choice' form, beyond standard EPR correlations. It is shown that the erroneous erasure claims arise from assuming that the improper mixed state of the signal photon physically prefers either the 'which way' or 'both ways' basis, when no such preference is warranted. The latter point is illustrated through comparison of the QE spatial state space with the spin-1/2 space of particles in the EPR-spin experiment.

## I. THE 'QUANTUM ERASER' EXPERIMENT: INTRODUCTION

The so-called 'quantum eraser' experiment ('QE'), first introduced in Kim et al (2000), involves a pair of entangled quanta (usually photons). One of these, called the signal (or 'system') photon *s*, is directed to a two-slit or two-arm apparatus and the other, the idler (or 'environment') photon *p*, is directed to a set of detectors equipped for various measurement options (or is subjected to polarization manipulations before detection). The 'delayed' version of the experiment consists of detecting *s* prior to *p*.

An example of the sort of claim analyzed and critiqued herein is:

"The which-path or both-path information of a quantum can be erased or marked by its entangled twin even after the registration of the quantum." (Kim et al, 2000.)

The main problem with this formulation of the experiment in terms of 'erasure of information' is that neither quantum has which-path nor both-paths information, so that it never has any relevant information to erase. (On the other hand, if its registration does yield such information, that information is never erased; discussion of this point is included in Sections III and IV.) While most physicists probably realize this, the usage of the terminology 'erasure of information' or 'marking of information' in connection with such experiments has given rise to enormous confusion and fallacious conclusions on the part of popular science writers and the public. In order to clearly see how the primary fallacy arises and to avoid falling victim to it, in what follows we will compare the state space of the 'QE' with the state space of the usual EPR-spin experiment with spin-1/2 quanta such as electrons.

## II. THE QUANTUM ERASER IS JUST LIKE THE EPR-SPIN EXPERIMENT

Note that the Hilbert Space description of the 'which path' and 'both-path' observables is isomorphic in key respects to the spin-1/2 observables 'Spin along **z**' and 'Spin along **x**'. That is, the 'which-slit' states 'slit A' and 'slit B' can be identified with the $\sigma_z$ basis:

$$\begin{aligned} &\text{"Slit A"} \leftrightarrow |z\uparrow\rangle = \begin{pmatrix} 1 \\ 0 \end{pmatrix} \\ &\text{"Slit B"} \leftrightarrow |z\downarrow\rangle = \begin{pmatrix} 0 \\ 1 \end{pmatrix} \end{aligned} \qquad (1)$$

Meanwhile, the 'both slits' states, commonly termed 'fringe' and 'antifringe,' can be identified with the **x**-spin or $\sigma_x$ basis:

$$\text{"Fringe"} \leftrightarrow |x\uparrow\rangle = \frac{1}{\sqrt{2}}\begin{pmatrix} 1 \\ 1 \end{pmatrix}$$
$$\text{"Antifringe"} \leftrightarrow |x\downarrow\rangle = \frac{1}{\sqrt{2}}\begin{pmatrix} 1 \\ -1 \end{pmatrix} \qquad (2)$$

The states in (2) are called 'symmetric' and 'antisymmetric' in Walborn et al (2002), who describe a version of the QE involving correlations of the spatial degree of freedom with photon polarization, such that measurement choices can be carried out by manipulation of the idler photon's polarization. While the overall state space of that version is more complicated, the essentials are nevertheless captured by the current analysis.

In view of the correspondence noted above, let us call the 'which slit' observable whose eigenstates are illustrated in (1), 'Z', and let us call the 'both slits' observable whose eigenstates are illustrated in (2), 'X'. Thus, the photon 'which slit' A and B states are $|Z\uparrow\rangle$ and $|Z\downarrow\rangle$ ; and the photon 'both slit' states, 'fringe' and 'antifringe,' are $|X\uparrow\rangle$ and $|X\downarrow\rangle$.

For convenience later, we'll note here that the projection operators corresponding to the above states are as follows:

$$P_{Z\uparrow} = \begin{pmatrix} 1 \\ 0 \end{pmatrix}\begin{pmatrix} 1 & 0 \end{pmatrix} = \begin{pmatrix} 1 & 0 \\ 0 & 0 \end{pmatrix};$$
$$P_{Z\downarrow} = \begin{pmatrix} 0 \\ 1 \end{pmatrix}\begin{pmatrix} 0 & 1 \end{pmatrix} = \begin{pmatrix} 0 & 0 \\ 0 & 1 \end{pmatrix} \qquad (3)$$

$$P_{X\uparrow} = \frac{1}{2}\begin{pmatrix} 1 \\ 1 \end{pmatrix}\begin{pmatrix} 1 & 1 \end{pmatrix} = \frac{1}{2}\begin{pmatrix} 1 & 1 \\ 1 & 1 \end{pmatrix};$$
$$P_{X\downarrow} = \frac{1}{2}\begin{pmatrix} 1 \\ -1 \end{pmatrix}\begin{pmatrix} 1 & -1 \end{pmatrix} = \frac{1}{2}\begin{pmatrix} 1 & -1 \\ -1 & 1 \end{pmatrix} \qquad (4)$$

In the QE experiment, a pair of correlated photons is created such that their spatial degree of freedom is an entangled EPR-type state, e.g.:

$$|\Psi\rangle = \frac{1}{\sqrt{2}}\left[A_1A_2 + B_1B_2\right] \tag{5}$$

where the subscripts denote the two photons, and here we have represented the state in the 'which slit' or Z basis. But let us not make the common mistake of thinking, based on our ability to write (5) in terms of A and B (i.e., $|Z\uparrow\rangle$ and $|Z\downarrow\rangle$), that 'photon 2 carries which-way (Z) information about photon 1' as opposed to both-ways information, since we can write (5) equally well in the both-ways basis.

To make explicit the conceptual equivalence of this state with the usual EPR spin situation, let us represent (5) using the equivalence discussed above, first using equation (1):

$$|\Psi\rangle = \frac{1}{\sqrt{2}}\left[\begin{pmatrix}1\\0\end{pmatrix}_1\begin{pmatrix}1\\0\end{pmatrix}_2 + \begin{pmatrix}0\\1\end{pmatrix}_1\begin{pmatrix}0\\1\end{pmatrix}_2\right] = \frac{1}{\sqrt{2}}\left[|Z\uparrow\rangle_1|Z\uparrow\rangle_2 + |Z\downarrow\rangle_1|Z\downarrow\rangle_2\right] \tag{6}$$

However, as noted above, the state $|\Psi\rangle$ can just as legitimately be written in the 'both slits' or X basis, via (2), i.e.:

$$|\Psi\rangle = \frac{1}{\sqrt{2}}\left[\frac{1}{2}\begin{pmatrix}1\\1\end{pmatrix}_1\begin{pmatrix}1\\1\end{pmatrix}_2 + \frac{1}{2}\begin{pmatrix}1\\-1\end{pmatrix}_1\begin{pmatrix}1\\-1\end{pmatrix}_2\right] = \frac{1}{\sqrt{2}}\left[|X\uparrow\rangle_1|X\uparrow\rangle_2 + |X\downarrow\rangle_1|X\downarrow\rangle_2\right] \tag{7}$$

This is nothing more than the observation in connection with the spin-EPR experiment that the state $|\Psi\rangle$ can be written as

$$|\Psi\rangle = \frac{1}{\sqrt{2}}\left[|\uparrow\rangle_1|\uparrow\rangle_2 + |\downarrow\rangle_1|\downarrow\rangle_2\right] \qquad (8)$$

*without regard to spin direction*. That is, the 'up' arrow can just as easily represent 'up along x' as 'up along z'; and similarly, the 'up' arrow can just as easily represent 'fringe pattern' as 'Slit A distribution'. The state $|\Psi\rangle$ has no preference for either basis, and its arbitrary representation with respect to one spin direction or spatial basis ('which slit' or 'both slits' basis) does not indicate that it has any intrinsic information about that spin direction or spatial basis. Put differently, Nature's informational content does not increase or decrease depending on whether humans elect to write down a state such as $|\Psi\rangle$ in a particular basis.

Now, let us see what information regarding the signal photon (if any) can be associated with the state that is used in the 'QE'. Given the entangled state $|\Psi\rangle$, neither photon is in a pure state independently of detection—i.e., it cannot be represented by a vector in Hilbert space. Each is in an improper mixed state, which cannot be written as a pure state or as a well-defined sum of pure state density operators (such as $|\psi\rangle\langle\psi|$)[1]. In particular, the density operator $\rho_1$ of the signal photon is just ½ times the identity, i.e.,

$$\begin{aligned}\rho_1 &= \frac{1}{2}\begin{bmatrix} 1 & 0 \\ 0 & 1 \end{bmatrix} = \frac{1}{2}\left(\begin{bmatrix} 1 & 0 \\ 0 & 0 \end{bmatrix} + \begin{bmatrix} 0 & 0 \\ 0 & 1 \end{bmatrix}\right) = \frac{1}{2}\left(\frac{1}{2}\begin{bmatrix} 1 & 1 \\ 1 & 1 \end{bmatrix} + \frac{1}{2}\begin{bmatrix} 1 & -1 \\ -1 & 1 \end{bmatrix}\right) \\ &= \frac{1}{2}\left(P_{Z\uparrow} + P_{Z\downarrow}\right) = \frac{1}{2}\left(P_{X\uparrow} + P_{X\downarrow}\right) = \frac{1}{2}\left(P_{\Theta\uparrow} + P_{\Theta\downarrow}\right).\end{aligned} \qquad (9)$$

In (9), we make explicit that the identity can be decomposed as the sum of the projection operators for the Z-outcomes (corresponding to slit A and slit B respectively); or, equally legitimately, as the sum of the projection operators for the X-outcomes

[1] That is, it cannot be considered an epistemic mixed state, in which the system is 'really' in one of the pure states and the sum represents ignorance of that state.

(corresponding to fringe and antifringe respectively); or as any other combination of the spatial modes, in analogy with some spin angle θ (or angles θ,ϕ for arbitrary spin directions in 3-space). Thus, the signal photon's state *contains neither 'which slit' information nor 'both slits' information*. In particular, (9) demonstrates that if the signal photon (encountering two slits A and B) is detected first at a screen measuring transverse distance *d*, the distribution exhibited, which is just noise, cannot be interpreted preferentially as a sum of the slit A and slit B patterns, since *exactly the same noise distribution is yielded as a sum of the fringe and antifringe patterns*.

This last point is crucial to avoid the common fallacy attending the 'QE' that when interference is absent, the which-way (Z) basis is singled out as physically or informationally applicable. That is not correct. The absence of interference in the case of a two-photon correlated state arises for a completely different reason than does the absence of interference in the case of a single-photon pure state. In the single quantum (pure state) case, under loss of interference the which-way (Z) basis is indeed singled out. But it is important not to confuse these two very different situations. As shown in (9), the absence of interference in the detection pattern of a component photon of an entangled state like (8) does not single out any basis. It is a reflection of the fact that each component photon is in an improper mixed state, with no preference for any basis.[2]

As noted above, the QE can be performed with a two-slit screen, in which case the detections of the signal photon record its horizontal position *d* after passage through the slits; or, as in Ma *et al* (2012), it can be performed with an interferometer, with detection at one of two detectors placed beyond a recombining half-silvered mirror. In the latter case, a phase difference of zero between the arms of the interferometer yields measurement of the 'both ways' observable Y (in analogy with spin along **y** for spin-1/2), with possible outcomes 'fringe' or 'antifringe'. Varying the phase difference corresponds to varying the azimuthal angle $\phi$ in the EPR-spin experiment. Signal photons prepared as components of the state (5) (equivalently ((6), (7), (8)) will be detected half the time at the 'fringe' detector and half the time at the 'antifringe' detector. Removal of the recombining mirror constitutes measurement of the 'which way' observable Z, and detections will again be evenly distributed. The same uniform noise distribution will apply for any phase difference between the arms (with recombining mirror in place). This again illustrates the ambiguity of the improper mixed state represented by the identity (9); it contains no information about any single-system observable.[3]

[2] In Walborn et al (2002), the signal photon's polarization may be detected at the screen, which constitutes a measurement of either the X or Z observable (depending on which polarization measurement takes place). Then it will project its partner into a state corresponding to that outcome, and the appropriate statistics will be found upon taking subsets of the coincidence count. If the Z observable is always measured via polarization at the screen, then the Z basis is preferred at $t_1$, and mutatis mutandis for the X observable. But such a preference does not arise based on a representation of the entangled state (8) in any particular basis. It can only arise through a physical measurement process, e.g. detection of the photon's circular polarization through transfer of angular momentum. After that outcome is registered, the attendant information is never 'erased' by a noncommuting measurement of its partner, as shown in subsequent sections.

## III. NO ERASURE, DELAYED OR OTHERWISE, IN THE 'QUANTUM ERASER' AFTER DETECTION

It should now be clear that when prepared in the state (8) (no matter what basis in which it might be written), the signal photon has no information about any particular observable to which it alone would be subject to measurement. What about after registration of the signal photon? A common assumption in discussions of the QE is that the signal photon's registration outcome represents either 'which way' or 'both ways' information that can later be 'erased' through a suitable measurement of its idler partner. In this section we will see, with the aid of the state-space equivalence to the EPR-spin situation, why it is inappropriate to think that detection at some point *d* (for the case of a moveable detector), or at a particular detector in an interferometer*,* represents information that can later be 'erased.' In the case of detection at *d*, that detection yielded neither 'which slit' nor 'both slits' information; it simply yielded *d* information that is not erased. Meanwhile, in the interferometer (or Walborn *et al*, 2002) version, if the signal photon's outcome corresponds to an eigenvalue either of the which-way or both-ways observable, such information is never erased just because its partner was measured in a different basis. Rather, it becomes part of an appropriate distribution dictated by the Born Rule. With the aid of the EPR-spin analogy, we can gain more insight into the situation.

---

[3] This is a well-established (but often forgotten) fact about component subsystems of non-separable states. See, for example, R.I.G. Hughes (1992).

Consider again the EPR-spin experiment with two electrons in the state (8). Each electron can be subjected to a measurement of spin in any desired direction. Suppose electron 1 is detected prior to electron 2. We don't normally call the component electrons either 'signal' or 'idler' (in fact this is one of the features that contributes to confusion in the QE). But to help reveal the fallacy attending the QE, let us consider electron 1 the 'signal' electron, analogous to the 'signal photon.'

Now, suppose we measure electron 1 along the $\theta$ axis (where the z axis is at an angle of zero). Let $\theta$ be some nonzero angle between the directions x and z. This is analogous to the measurement of the signal photon at some position *d* along the screen, except that there are only two possible outcomes for the spin measurement, corresponding to the eigenstates $|\theta\uparrow\rangle$ and $|\theta\downarrow\rangle$. (This is a minor difference that does not undermine the analogy, and in fact it becomes an exact analogy for the interferometer case, for the azimuthal angle $\phi$, to be discussed later.) So suppose our electron 1 ('signal electron') is detected in the state $|\theta\uparrow\rangle$. Its registration then yields $\sigma_\theta$ information, *but nothing else*! For example, nobody would say, based on the ability to write down the state (8) in the $\sigma_z$ -basis, (as in (6)) that the outcome $|\theta\uparrow\rangle$ has anything to do with '$\sigma_z$ information'; nor would they say, based on the ability to write (8) in the x-basis (as in (7)), that the same outcome has anything to do with '$\sigma_x$ information.' The fact that the correlation can be expressed in a particular basis does not warrant the idea that either electron is 'marked' with 'information' pertaining to the basis in which someone chose to represent (8). So, similarly, a photon's registration at position *d* on the screen provides *only D-observable* information, and nothing else. It does not represent either 'which slit' or 'both slits' information.

Crucially, however, the first electron's detection in the state $|\theta\uparrow\rangle$ *does project its partner, electron 2, into the pure state* $|\theta\uparrow\rangle$, based on the state (8). That is standard EPR correlation. If we later measure the spin of electron 2 along the **z** axis, we will get $\sigma_z$ outcomes distributed according to the Born Rule applying to the spin-½ Hilbert space, i.e.:

$$\mathrm{Prob}\left(z\uparrow|\,\theta\uparrow\right)=\left\langle z\uparrow\middle|\theta\uparrow\right\rangle\left\langle\theta\uparrow\middle|z\uparrow\right\rangle=\cos^2\frac{\theta}{2};\quad \mathrm{Pr}\,ob\left(z\downarrow|\,\theta\uparrow\right)=\sin^2\frac{\theta}{2} \qquad (10)$$

where in the first expression in (10) we explicitly show the symmetry of the Born probabilities with respect to conditionalization. For convenience, we also write down the relevant Born probabilities for the case in which electron 1 is first detected in the state $\left|\theta\downarrow\right\rangle$:

$$\mathrm{Prob}\left(z\uparrow|\,\theta\downarrow\right)=\sin^2\frac{\theta}{2};\quad \mathrm{Pr}\,ob\left(z\downarrow|\,\theta\downarrow\right)=\cos^2\frac{\theta}{2} \qquad (11)$$

Similarly, the signal photon's detection in the screen pixel state $\left|d\right\rangle$ projects its idler partner into a pure state that dictates specific probabilities for its detection in whichever basis it is measured, whether Z ('which way') or X ('both ways'). For example, suppose the signal photon is detected at a value of *d, call it* $d_{df}$, corresponding to a dark spot in the 'fringe' interference pattern. (Recall that the state corresponding to 'fringe' is $\left|X\uparrow\right\rangle$.) This means that

$$\mathrm{Prob}\left(X\uparrow|\,d_{df}\right)=\left\langle X\uparrow\middle|d_{df}\right\rangle\left\langle d_{df}\middle|X\uparrow\right\rangle=0 \qquad (12)$$

and *its idler partner will be forbidden*, based on its projected pure state resulting from its signal partner's detection, from being found in the state $\left|X\uparrow\right\rangle$. (The correlation is enforced through the photons' momentum entanglement.) So naturally, when the coincidence count is sorted, that particular signal photon's registration at $d_{df}$ will never appear in the distribution for the 'fringe' state $\left|X\uparrow\right\rangle$. Nothing was 'erased' through its idler partner's detection; rather, the signal photon's detection at $d_{df}$ steered its partner away from one state and towards others. Its $d_{df}$ outcome has a chance of appearing in the idler distributions for $|X\downarrow\rangle, |Z\uparrow\rangle$, and $|Z\downarrow\rangle$, not through any 'erasure,' but simply because its detection has prepared its partner in a state with a nonzero projection on any of those states. The same basic principle holds for any value of $d$: as in (12), the probabilities for idler outcomes will be conditionalized on the $d$ outcome of the signal photon.

Thus, *the signal photon's d-outcome steers the idler*, and that is what enforces the correlations that are observed through the coincidence count.[4] The fact that there is no legitimate basis for a retrocausal 'delayed choice' effect involved in these correlations is also seen clearly by comparison with the EPR-spin case, as follows. Suppose we look at a coincidence count of the electrons in the EPR-spin case, choosing the subset of electron 2's ('idler') outcomes $|z\uparrow\rangle$. We will find their electron 1 ('signal') partners' $\theta$-axis outcomes (analogous to the photon $d$-outcomes) appropriately distributed according to the probabilities (10), which are symmetric in the conditionalization; i.e., $\text{Prob}(z\uparrow|\theta\uparrow) = \text{Prob}(\theta\uparrow|z\uparrow)$, etc.

Nobody concludes, based on the observation that the $\theta$-outcomes are distributed according to the Born Rule (10),(11) that electron 1's 'x-axis information has been erased by delayed measurement of electron 2 along the z axis' (or vice versa), nor should they; because *electron 1 never had any x- nor z-related information*. All it ever had was $\theta$-information, upon registration. A distribution of such outcomes corresponding to the probabilities (10) and (11), enforced by electron 1's steering of its partner electron 2, is never taken as an indication that electron 2's detection 'erased' any information about **z** or **x** possessed by the electron 1. Nor should it be. And of course, if electron 1 is measured with respect to **z** (rather than $\theta$), its outcome yields **z** -information that is not erased, no matter the basis with respect to which its partner is measured.

[4] While this mutual 'steering' is counterintuitive, as analyzed and discussed by Schrodinger (1936), again, it is no different from what goes on in the standard EPR situation, which does not involve 'erasure' or explicit retrocausation—just the usual fact that entangled quantum systems exert apparently nonlocal influences on one another.

## IV. INTERFEROMETER VERSION OF THE 'QE'

Ma et al (2013) present an interferometer version of the so-called quantum eraser involving correlation of the polarization degree of freedom of the idler ('environment') photon with the spatial degree of freedom of the signal ('system') photon. In this case, we have only two measurement outcomes for the signal photon, corresponding to the two detectors for the particular setting of the interferometer. Without the final recombining beam splitter, the two detectors measure the observable Z; with the final recombining beam splitter, observables corresponding to Y, or other arbitrary 'angles' $\phi$ (corresponding to varying phase difference between the arms) are measured. Again, we have nothing more than standard EPR correlations, which we can verify by comparison to the EPR-spin experiment.

The two photons are prepared in a correlated state analogous to (5), with the polarization degree of freedom of the idler photon correlated to the spatial degree of freedom of the signal photon. However, as we have seen in the previous sections, such a state does not imply information about any property of the signal photon (such as 'which way'), since (5) can be represented in an arbitrary basis with respect to polarization and spatial state, as reflected in its more general form (8). The signal photon is no more 'marked' for any particular measurement basis than are either of the electrons in the state (8).

Specifically, the fact that horizontal polarization of the idler is correlated to the signal photon's spatial state A, while vertical polarization of the idler is correlated to the signal photon's spatial state B, does not constitute the availability of information about the 'which way' (Z) status of the signal photon. The authors (perhaps inadvertently) perpetuate this fallacy when they say, based on their equation (1) showing this correlation in the Z basis:

> "The environment photon thus carries welcher-weg information about the system photon."

However, the environment photon has no specific which-way information about the system photon, because the entangled state is oblivious to basis. This statement is no more correct than saying of two entangled EPR-spin electrons in state (8) that ‘electron 2 carries z-spin information about electron 1’. That is misleading in the sense that it implies a privileged status for the $\sigma_z$ basis that has no physical warrant, and utterances like these are avoided in the EPR-spin experiment, for good reason. The only ‘information’ electron 2 contains is that its spin outcome will be the same as the spin outcome of electron 1 for *any* spin direction—corresponding to *any* spatial basis (X or θ or Z) for the signal photon.

The correlated photon state can equally well be written in the both ways (in this case, Y) basis, in which case the circular polarization states R and L of the idler correlate to the ‘fringe’ and ‘antifringe’ states of the signal photon. In fact the authors note this point, saying that both representations are ‘equally fundamental,’ but they omit the implication that this complete lack of preference of the state (8) for any relevant single-system observable means that the environment photon carries no more information about the system photon for any one observable than another—*and that is why no information pertaining to any particular basis is ever ‘erased.’*

Again, both photons are in an improper mixed state—the identity—without preference for any observable; thus, nothing at all about the signal photon’s spatial state properties is ‘marked’ subject to later ‘erasure,’ any more than an electron’s $\sigma_z$ status is marked for ‘erasure’ in a state like (8). A choice is made whether to measure the equivalent of Y (via circular polarization) or Z (via linear polarization) on the idler photon, and the outcomes of the signal photon in either Y or Z, or some arbitrary phase angle ϕ, exhibit the required correlations upon coincidence counting. The correlations found are not due to any ‘erasure’ but simply due to mutual enforcement of the Born Rule, just as in the EPR-spin case. The two photons steer one another, just as two spacelike-separated electrons in the EPR-spin experiment undergo mutual steering. For example, in the EPR-spin case, ‘signal’ electrons that are measured along **z** and registered in an outcome of the $\sigma_z$ observable only convey information about $\sigma_z$, and no other observable. Measurement of their partner with respect to a noncommuting observable $\sigma_\phi$ or $\sigma_y$ does not constitute 'erasure' of any $\sigma_z$ information, and quite rightly, nobody says that about the equivalent correlations observed in the EPR-spin experiment. The coincidence count simply provides the appropriate distribution of $\sigma_z$ outcomes, which through EPR correlation conform to the probabilities (10) and (11). Nothing more than this goes on in the 'QE' experiment.

An anonymous reviewer puts it very well regarding popular confusion about experiments like that of Ma et al, resulting from the presentation of these experiments in terms of 'erasure': "The misleading terminology leads science writers and journalists to misunderstand the experiment and make claims that what you do on Tenerife is able to cause what happens on La Palma, even somehow after it happens. That's clearly not the case."

## V. CONCLUSION

It has been demonstrated that the observed correlations of the 'quantum eraser' and 'delayed choice quantum eraser' are fully accounted for by standard EPR correlations; there is no necessary 'temporal nonlocality' obtaining in the QE experiment, beyond the usual fact that spacelike-separated detections have no absolute temporal order.[5] The habitual, but inappropriate, characterization of the results in these experiments as 'erasure' is a fallacy resulting from overlooking any or all of the following:

(i) the fact that the signal photon has no intrinsic information about either the 'which way' or 'both ways' observable based on its prepared state. (Improper mixed state)

[5] If the detections are timelike separated, then the quantum detected first unambiguously projects the second quantum into a pure state, which dictates the probabilities of its outcomes for all measurements, and again, there is no 'delayed' or explicitly retrocausal effect.

(ii) the requirement for coincidence counting and statistical analysis of the data to sort the *d* detections into the correct sub-ensembles. (Lack of interference at screen does not privilege the which-way basis)[6]

(ii) the fact that the signal photon detections project their idler partners into pure states whose statistical properties, upon measurement, will correctly reflect the *d* value of their partner signal photon's detection. (EPR steering)

In the case of interferometer measurements, or the two-slit experiment with polarization entanglement (as in Walborn *et al*, 2002), each of the signal photons' outcomes provide information only about the actual observable measured. So, for example, if signal photons are first subject to a 'which way' (Z) measurement, their registration outcome provides information about the Z observable, and their correlated partner is projected into a pure state, either $|Z\uparrow\rangle$ or $|Z\downarrow\rangle$, based on that outcome. Then the idler partner's outcomes are distributed in accordance with the Born Rule, conditionalized on its projected pure state resulting from the signal photon's detection. If a 'both ways' (X or Y) measurement performed on the idler partner, that does not 'erase' its signal photon's Z outcome or information. Rather, the signal photons whose idlers ended up i a 'fringe' state, i.e., $|Y\uparrow\rangle$, will simply be found to be evenly distributed over $|Z\uparrow\rangle$ and $|Z\downarrow\rangle$. This is no different from the EPR-spin case, where instances of 'electron 2' detected in the state $|y\uparrow\rangle$ will have their 'electron 1' partners' outcomes evenly distributed over $|z\uparrow\rangle$ and $|z\downarrow\rangle$. Nobody invokes 'erasure' or 'delayed erasure of z-spin information' in the latter case; and indeed no such information was erased. We just have EPR correlations, and the situations are completely isomorphic.

It is curious that the very same sorts of correlations have been known to arise in the EPR-spin experiment, yet they have never been misconstrued as 'erasure of information,' delayed or otherwise, with respect to any particular spin-axis observable. The stubborn

---

[6] This applies to the version of the experiment without polarization entanglement. If polarization entanglement yields a which-way or both-ways basis preference at the signal detection screen, we still have ordinary EPR 'steering,' as discussed below.

'erasure' concept that has attached itself to experiments involving 'which way' or 'both ways' properties may be due to the fact that these spatial properties directly affect our perceptions by creating visual patterns. We then identify the concept of 'information about an observable' with these patterns. But they are simply Born Rule distributions of outcomes—i.e., exemplars of conditional probabilities. As we have seen, in an entangled state expressible as (5), (6), and (8), neither quantum has information about any particular observable apart from its registration in an outcome of the specific measurement performed on it, which is never erased.[7] The patterns seen only after coincidence count comparisons are nothing more than conditional Born probability distributions, and nothing at all is erased in order for these to arise, any more than any electron's spin-measurement outcome or information is erased in the EPR-spin experiment. The construal of such correlations as involving 'erasure' is based on an illusion, and physics is about dispelling illusion. Even if the physicists who conduct these experiments understand the fact that the component subsystems carry no basis-specific information about their partners, the terminology 'erasure of information' (or 'marking of information') is distinctly inappropriate given their improper mixed states, and has given rise to enormous confusion. It is time to let go of the misleading misnomer 'quantum eraser' for these spatial-space EPR experiments.

Acknowledgments.
The author would like to thank David Ellerman and an anonymous referee for valuable comments.

[7] David Ellerman (2015) makes a similar point in his critique of the usual conclusions regarding the 'delayed choice quantum eraser.'